\documentclass[twocolumn]{article}

\usepackage{graphicx}
\usepackage[left=2cm,right=2cm,top=2cm,bottom=2cm]{geometry}
\usepackage{amsmath}
\usepackage{amsfonts}
\usepackage{amssymb}
\usepackage{cite}
\usepackage{hyperref}

\begin{document}

\title{Experimental characterization of transitions between locking regimes in a laser system with weak periodic forcing}
\author{J Tiana-Alsina, C Quintero-Quiroz, M. C. Torrent, C Masoller.
\\
Universitat Polit\`ecnica de Catalunya, Departament de F\'isica,\\ Sant Nebridi 22, 08222 Terrassa, Barcelona, Spain
}%

\date{\today}

\twocolumn[
\maketitle
The entrainment (or locking) phenomenon, by which an oscillator adapts its natural rhythm to an external periodic signal, is well-known in physics, chemistry, biology, etc.; however, controlling an stochastic nonlinear system with a small-amplitude signal is a challenging task, and systems that allow for low-cost experiments are scarce. Here we use a semiconductor laser with optical feedback, operated in the regime where it randomly emits abrupt spikes. We quantify the quality of the entrainment of the optical spikes to periodic, small-amplitude electric perturbations of the laser pump current. We use the success rate (SR) that counts the number of spikes that occur within a short time window after each perturbation, and the false positive rate (FPR) that counts the additional spikes that occur outside the window. The ROC curves (SR vs. FPR plots) uncover parameter regions where the electric perturbations fully control the laser spikes, entraining them, such that the laser emits, shortly after each perturbation, one and only one spike (i.e., SR=1 and FPR=0). We also characterize the locking-unlocking transitions when the perturbation amplitude and frequency vary. 
\vspace{1cm}
]
\vspace{1cm}


\section{Introduction}
Many natural systems show the ability to adjust their natural rhythms to follow an external periodic signal~\cite{winfree2001geometry,pikovsky2003synchronization,kuramoto_book_2003}. The phenomenon, known as entrainment or locking, has been observed in lasers~\cite{elsasser_prl_1989,erneux_1997,sukow2000entraining,mindlin_pre_2001,Barland_PRE:2013}, chemical reactions~\cite{kiss_2013,kiss_nat_comm_2016}, neuronal oscillations~\cite{andre_dante_1998,neurons_2008}, cicardian cells~\cite{glass_1982,circardian_1996,circardian_2001,circardian_2018}, etc. 

Entrainment is typically achieved by increasing the amplitude of the forcing signal until the system adjusts its frequency to that of the signal. However, strong forcing might damage the system 
and control methods that apply weak periodic perturbations are often needed. 
A very popular control approach, proposed by Ott, Grebogi and Yorke \cite{PhysRevLett.64.1196}, consists of forcing a system to adjust its natural rhythm through small periodic perturbations that stabilize an unstable periodic orbit of the system. This technique is successful if the system has an unstable orbit that can be stabilized. However, many biological systems display noise-induced rhythms (while in the absence of noise, they rest in a steady-state), and it is important to test methods for entraining a noisy oscillator to a weak external signal. 

Noisy entrainment can be difficult to identify and to quantify, and appropriated measures are needed. 
A well-known tool is the phase-response or phase-resetting curve (PRC) that describes the effect of a perturbation in the phase of the oscillator \cite{prc_2005,netoff2012experimentally,mirasso_2017}. The PRC simplifies the description of complex, stochastic dynamics to a one-dimensional phase dynamics,  allowing to determine whether the phase of the oscillator is locked to the external signal, but it has the drawback that one needs to estimate the phase, which can be difficult when the oscillator is inherently noisy and/or when its dynamics involves different timescales. 

In this paper we study the entrainment of a noisy oscillator to a weak external signal using  receiver operating characteristic (ROC) curves. A ROC curve allows to quantify the diagnostic ability of a binary classifier as a function of its classification threshold. ROC curves, developed during World World II, are nowadays routinely used by machine learning algorithms for classification, but have not yet been employed, to the best of our knowledge, to quantify entrainment. 

The experimental system investigated is a semiconductor laser with optical feedback from an external reflector. It displays stochastic dynamics with a rich variety of nonlinear behaviors~\cite{ohtsubo2012semiconductor,mork1992chaos,sciamanna2015physics}. Here we focus on the so-called low-frequency fluctuations (LFF) regime, where the laser emits a spiking output: during a spike the intensity drops abruptly and then recovers gradually. The parameter region where the LFF regime occurs is quite wide, and includes a sub-region where the emitted spikes are consistent with events seeded by noise~\cite{srep_2016,chaos_2017}. In this region, we aim to control the spikes via periodic, small-amplitude electric perturbations of the laser pump current. 
\section{Experimental Setup}

The experimental setup was described in \cite{Sorrentino:15,tiana2018experimental}. Pulse-down periodic perturbations are applied to the laser current and 
the control parameters are the perturbation amplitude, $A_{mod}$, frequency, $f_{mod}$, and the dc value of the laser current, $I_{dc}$, which controls the natural frequency of the spikes, $f_0$. 

 

For each set of parameters a time series of the laser intensity with $N=10^7$ data points was recorded with $2$~GS/s sampling rate, which allowed to capture the intensity dynamics during 5~ms. 

\section{Results}

\begin{figure}[tb]
\begin{center}
\includegraphics[width=\columnwidth]{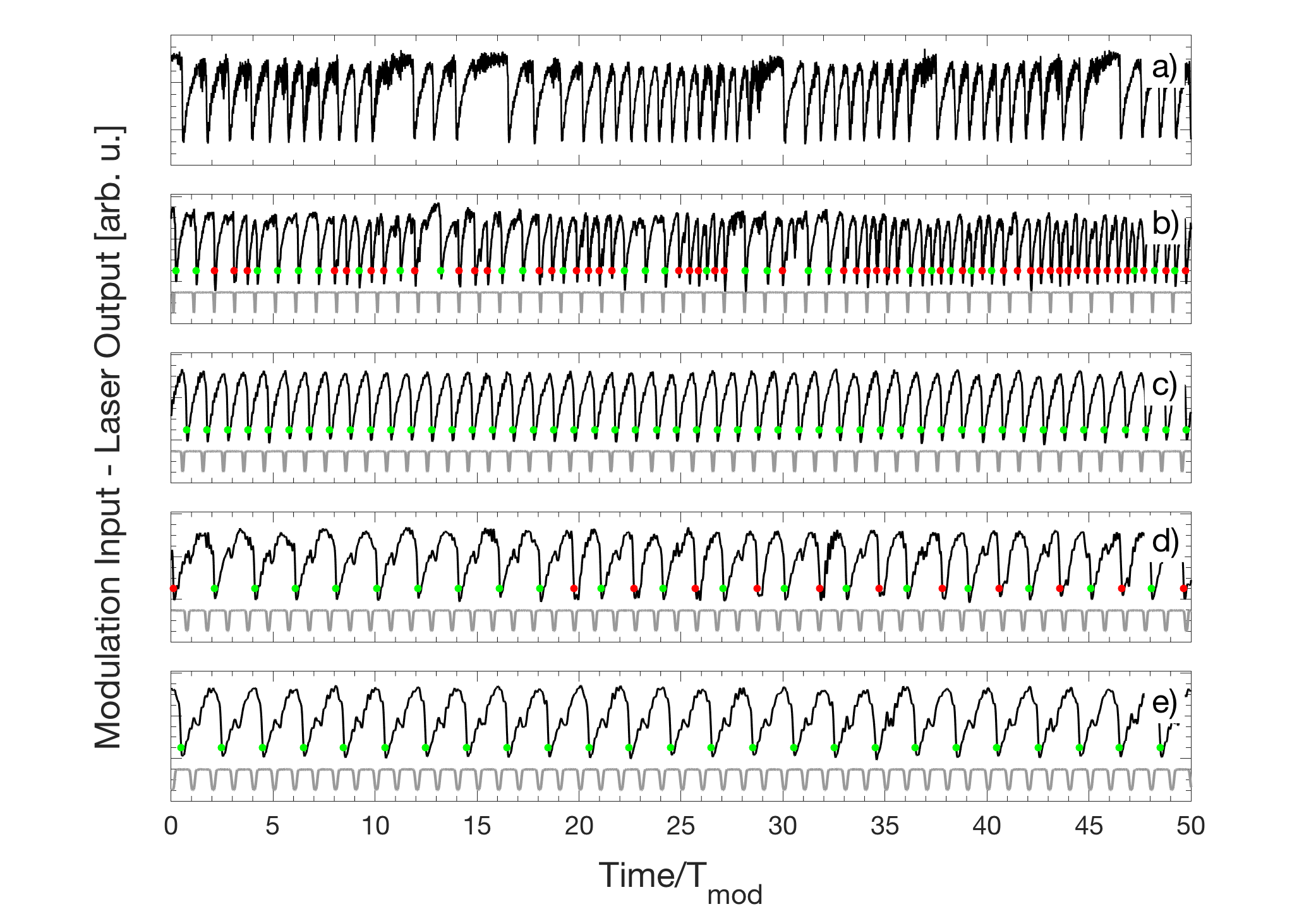}
\end{center}
\caption{Time series of the laser intensity (black line, normalized to zero mean and unit variance) and the pulse-down waveform applied to the dc pump current (gray lines, shifted vertically for clarity). Green dots represent the spikes that occur shortly after a perturbation, while red dots mark those spikes that are considered non-induced by a perturbation. The dc pump current is $I_{dc} = 27$~mA and the modulation amplitude is $2.3$\% of $I_{dc}$. Panel (a) shows the unforced dynamics ($A_{mod}=0$). For this pump current the natural frequency of the laser spikes is $f_0=15$~MHz and the horizontal axis is normalized to $1/f_0$. Panel (b) shows the intensity dynamics when $A_{mod}=$2.3\% of $I_{dc}$ the perturbation frequency, $f_{mod} = 10$~MHz. Panels (c), (e) display locking $1$:$1$, and $2$:$1$ for $f_{mod} = 20$~MHz, and $41$~MHz respectively; panel (d) shows the transition between locking $1$:$1$ and $2$:$1$ observed at $f_{mod} = 30$~MHz. In panels (b)-(e) the horizontal axis is normalized to $1/f_{mod}$.}
\label{fig:Time series}
\end{figure}

Figure~\ref{fig:Time series} displays typical examples of the intensity dynamics when the laser current is not perturbed (panel a), and when it is periodically perturbed (panels b-c) with $A_{mod}=$2.3\% of $I_{dc}$ and different perturbation frequencies. 

The spikes which occur shortly after a current perturbation are considered to be induced by the perturbation and are indicated with green dots (in the following, they will be referred to as \textit{true positives}). The other spikes will be referred to as \textit{false positives} and are indicated with red dots.
In Fig.~\ref{fig:Time series}(b) the frequency of the perturbations is lower than the natural frequency of the spikes ($f_{mod} = 10$~MHz and $f_0=15$~MHz). It is observed that after each perturbation the laser emits a spike, but in between perturbations the natural dynamics prevails and thus, most of the spikes are spontaneous (false positives). For a higher frequency, Fig.~\ref{fig:Time series}(c), locking $1$:$1$ is observed since every perturbation triggers a spike. For a higher frequency, Fig.~\ref{fig:Time series}(d), there is a transition between locking $1$:$1$ and $2$:$1$. In this region the spikes cannot follow the fast external perturbations and some spikes are delayed with respect to the perturbations. By further increasing the perturbation frequency, the spike rate adjusts such that there is one spike every two perturbations, Fig.~\ref{fig:Time series}(e).

\subsection{Inter Spike Intervals Distributions}

\begin{figure}[tb]
\begin{center}
\includegraphics[width=\columnwidth]{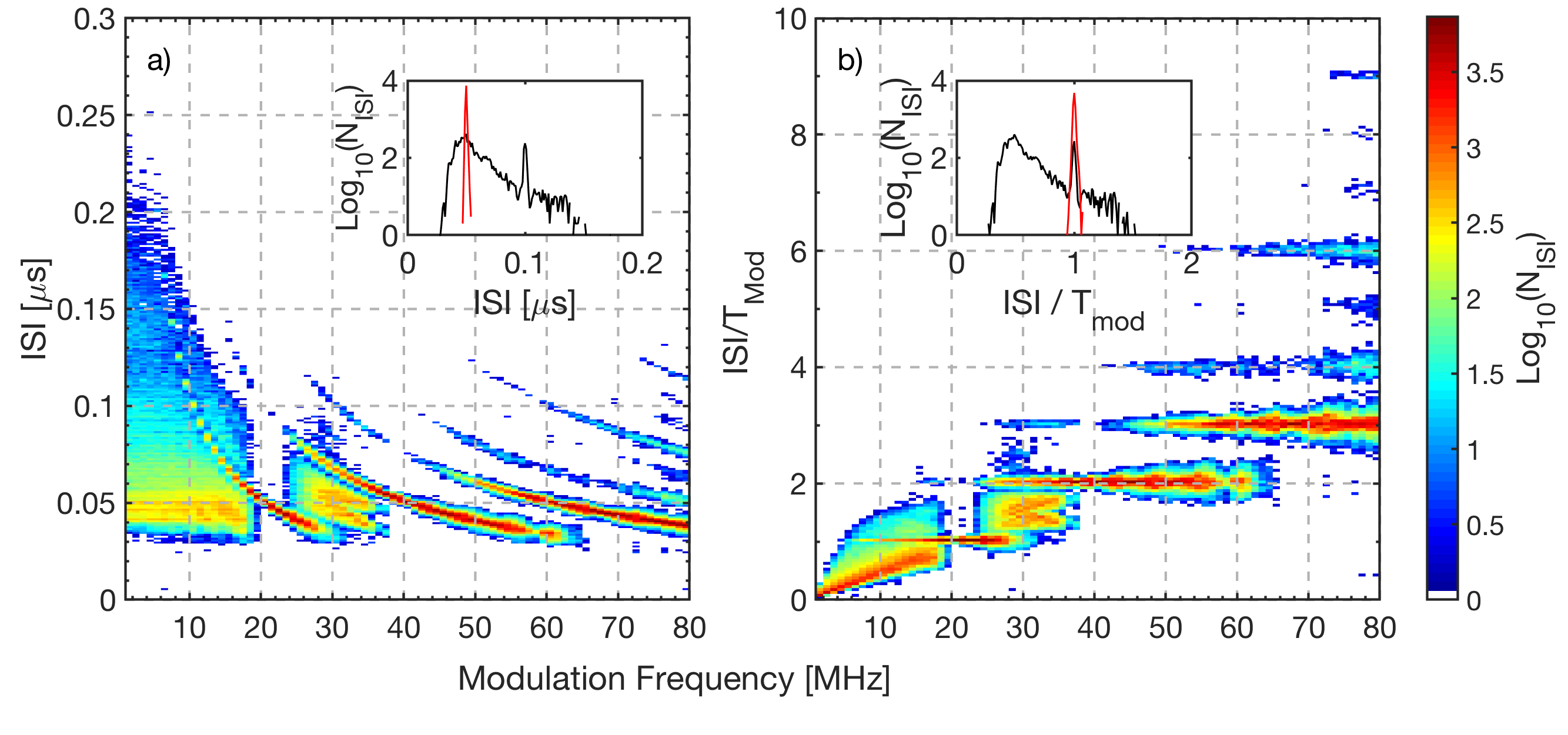}
\end{center}
\caption{Inter-spike interval (ISI) distribution as a function of the perturbation frequency for $I_{dc}=27$~mA and $A_{mod}=0.62$~mA (2.3\% with respect to $I_{dc}$). In order to enhance the plot contrast, the color scale indicates the logarithmic of the number of intervals (the white color stands for zero counts). In panel (a) the vertical axis is the ISI, while in (b), it is normalized by the perturbation period. The insets display typical ISI histograms when the laser is mainly driven by its natural dynamics (black line for $f_{mod}=10$~MHz) and when is driven by the external perturbations (red line for $f_{mod}=20$~MHz).}
\label{fig:ISI_hist}
\end{figure}

The variation of the spike rate with the frequency of the external signal is typical of the entrainment phenomenon. 
To analyze the entrainment quality we study the distribution of the time intervals between consecutive spikes (the inter-spike intervals, ISIs). We present in Fig.~\ref{fig:ISI_hist} the ISI distribution (in color code) vs. the perturbation frequency, keeping fixed $A_{mod}$ and $I_{dc}$. The ISI distribution is presented in two ways: in Fig.~\ref{fig:ISI_hist}(a) the vertical axis is the time interval between spikes (ISI), while in Fig.~\ref{fig:ISI_hist}(b), it is normalized to the modulation period (here the histograms are computed with bins centered at integer multiples of $T_{mod}=1/f_{mod}$). 

In Fig.~\ref{fig:ISI_hist}(a), as $f_{mod}$ increases we observe the transition from no locking to $1$:$1$ locking. At low frequencies (from $0$ to $15$~MHz) the laser behaves as if it is not driven by the external signal and the natural noisy dynamics dominates. This is revealed by a broad ISI distribution which has a narrow peak at $f_{mod}$ (an example of the ISI distribution in this frequency range is presented in the inset, black line). The peak represents the likelihood of two consecutive spikes being separated by a time interval equal to the period of the external signal. 

As the frequency increases, the spikes lock to the external signal and the ISI distribution becomes very narrow, as seen in the insets in Fig.~\ref{fig:ISI_hist}, red lines. In the locking regions the spikes are mainly controlled by the external perturbations, and the ISI distribution peaks at $nT_{mod}$, where $n$ is an integer number.

A large transition region characterized by a broad ISI distribution is observed between locking $1$:$1$ and $2$:$1$. It can be seen in Fig.~\ref{fig:Time series}(d) that this region the dynamics is characterized by a reorganization of the spikes, which no longer fit in one period (as in the $1$:$1$ region), but an interval of two periods is too long for a single spike (as in the $2$:$1$ region). Therefore, after a perturbation, some spikes occur before the next perturbation while others occur after the next perturbation. 

Another feature that can be observed in Fig.~\ref{fig:ISI_hist}(a) is that the spike rate cannot be too fast (the smallest ISI is about $0.03$~$\mu$s). When the perturbation frequency increases and the most probable ISI reaches this minimum time, the transition to the next locking curve starts. The minimum ISI (referred to as \textit{refractory time}) is due to the fact that after each spike, a step-like recovery occurs, and during the recovery process another spike is not emitted (except at the highest pump currents as will be discussed latter). 

In Fig.~\ref{fig:ISI_hist}(b) the $n T_{mod}=\langle ISI\rangle$ curves are converted into horizontal plateaus due to the normalization by $T_{mod}$. In this plot it is clearly observed that, as the modulation frequency increases, the ISIs become larger multiples of $T_{mod}$ as the laser spikes are spaced by an increasing number of perturbation cycles. At very high frequencies (above $~50$~MHz) the ISI distribution is not unimodal but has several peaks centered at $nT_{mod}$.  This normalized representation of the ISI distribution has the advantage that the locking regions are easy to identify, but the refractory time is not. Due to the normalization, the distribution of the natural spikes (which is independent of $f_{mod}$ in the non-normalized representation of the ISI distribution) is converted in a narrow tilted line and the broad nature of the ISI distribution at low frequencies is not visible. 

\begin{figure}[tb]
\begin{center}
\includegraphics[width=\columnwidth]{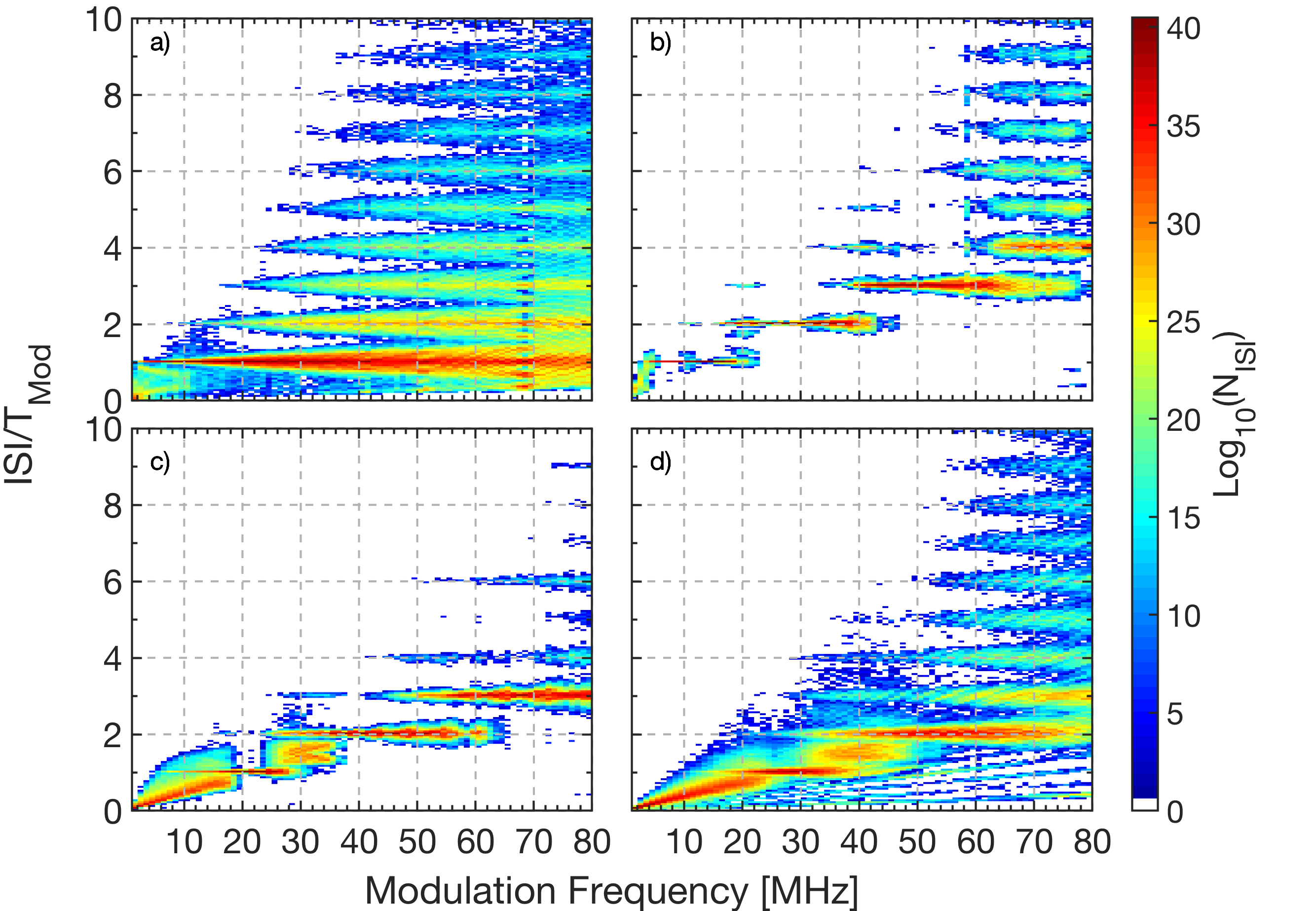}
\end{center}
\caption{Normalized ISI distributions as a function of the perturbation frequency for four dc values of the pump current: (a) $I_{dc}= 25$~mA,  (b) 26~mA, (c) 27~mA, and (d) 28~mA. The perturbation amplitude is $A_{mod}=0.62$~mA. The color code represents the logarithmic of the number of ISI counts. The white color indicates zero counts.}
\label{fig:ISI_hist_4panels}
\end{figure}

Figure~\ref{fig:ISI_hist_4panels} displays the ISI distribution for four dc values of the laser pump current. The lowest pump current, $I_{dc}= 25$~mA, corresponds to the onset of the LFF regime and the spikes are not really well defined. The laser is sensitive to the external perturbations, but the perturbations are not always able to trigger spikes. Therefore, we observe in Fig.~\ref{fig:ISI_hist_4panels}(a) that the ISI distribution has peaks at several multiples of the perturbation period. 

At higher pump currents [Figs.~\ref{fig:ISI_hist_4panels}(b) and (c) for $I_{dc}=$~26~mA and 27~mA respectively] the laser emits, without external perturbations, well defined spikes which can be entrained to the current perturbations, as revealed by well defined plateaus. For example, in panel (b) $1$:$1$ locking occurs for $f_{mod}$ in [5-13]~MHz, while in panel (c), $1$:$1$ locking occurs for $f_{mod}$ in [19--23]~MHz. The shift in the position of the locking regions with $I_{dc}$ is due to the fact that the natural spike frequency increases with the pump current~\cite{yanhua}, and therefore, for higher pump currents, higher frequencies are needed to entrain the spikes. 

With further increase of the laser current the coherence of the spikes is gradually lost. In this region is harder to entrain the spikes since the natural dynamics tends to dominate over the external perturbations. As it can be seen in Fig.~\ref{fig:ISI_hist_4panels}(d) for $I_{dc}= 28$~mA, for low perturbation frequency, the ISI distribution is similar to that in Figs.~\ref{fig:ISI_hist_4panels}(b,c), however, when the frequency is increased, no well-defined plateaus are seen. 
In Fig.~\ref{fig:ISI_hist_4panels}(d) we also note several tilted lines that correspond to ISIs shorter than the most probable ISI. The slope of these lines is an integer multiple of 5 ns, which is the feedback delay time. They are only seen at high pump currents and correspond to the detection of spikes during the recovery process of the previous spike (which occurs in steps whose duration is the feedback delay time). They are due to the lost of coherence of the individual spikes, which, at high currents, become too fast and irregular, signaling the onset of the so-called coherence collapse regime.


\subsection{ROC curves}

In order to quantify the degree of entrainment we use ROC curves, which are obtained by plotting the true positive rate (TPR, also referred to as success rate, SR) as a function of the false positive rate (FPR), for different values of the control parameters. The SR measures the response of the laser per perturbation cycle: if the laser emits one spike after each perturbation, SR$=1$, if it emits one spike every two perturbations, SR$={1}/{2}$, etc. Only spikes emitted within a detection window of duration $\tau$ are considered as spikes induced by the perturbation. The length of the window, $\tau=15$~ns, has been chosen such that only one spike can be emitted within this interval of time (see \cite{tiana2018experimental} for a detailed discussion). On the other hand, the FPR measures the spikes which are emitted outside this window. FPR$=0$ indicates that the spikes are always emitted within the time interval $\tau$ after a perturbation, while FPR$=1$ indicates that the laser does not emit any spike within this time interval. 

The TPR vs. FPR plots (ROC curves) allow identifying the optimal combination of experimental parameters ($A_{mod}$, $f_{mod}$, and $I_{dc}$) that produce the best entrainment: if we want to generate an optical spike for each electric perturbation, the optimal parameters are those that give points in the curve that are closest to the top-left corner (i.e., SR=1 and FPR =0).

\begin{figure}[tb]
\begin{center}
\includegraphics[width=\columnwidth]{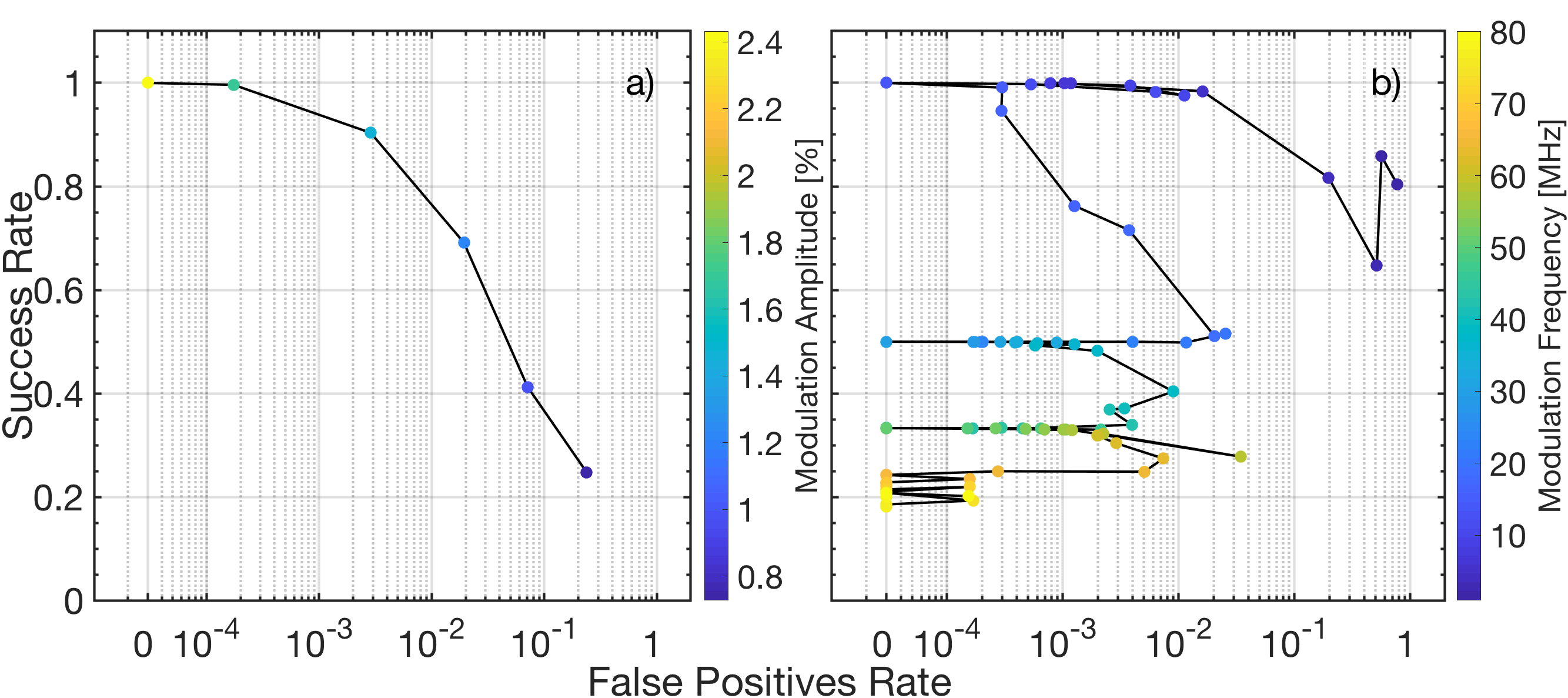}
\end{center}
\caption{ROC curve to track transition to locking when (a) the perturbation amplitude increases ({in color scale}) while keeping constant the frequency ({$f_{mod}=14$~MHz}) and when (b) the perturbation frequency increases ({in color scale}) while keeping constant the amplitude ({$A_{mod}=0.62$~mA}). The dc value of the pump current is $I_{dc}=26$~mA. To represent in logarithmic scale the value FPR$=0$, we have set it to $3\times 10^{-5}$ (labeled as 0 in the x-axis).}
\label{fig:ROC}
\end{figure}

Figure~\ref{fig:ROC}(a) displays the transition to locking $1$:$1$ when the perturbation amplitude is increased while the frequency is kept constant (iso-frequency line with $f_{mod}=14$~MHz). We note that we reach perfect 1:1 locking  (SR=1 and FPR =0) for $A_{mod}=2.4\%$. 
Figure~\ref{fig:ROC}(b) displays the transition as the perturbation frequency grows while the amplitude is kept constant (iso-amplitude line with $A_{mod}=2.4\%$). Here, the transition from SR$=1$ and FPR$=0$ to SR$={1}/{2}$ and FPR$=0$, as well as the transition from SR$={1}/{2}$ and FPR$=0$ to SR$={1}/{3}$ and FPR$=0$ are observed. During the transition from one locking regime to another, there is an increase of the number of false positives, followed by a decrease. The increase is due, as discussed before, to the re-organization of the spikes: they cannot follow the external signal as it becomes faster, but, on the other hand, two periods of the signal is a too long time interval for only one spike. 

\begin{figure}[tb]
\begin{center}
\includegraphics[width=\columnwidth]{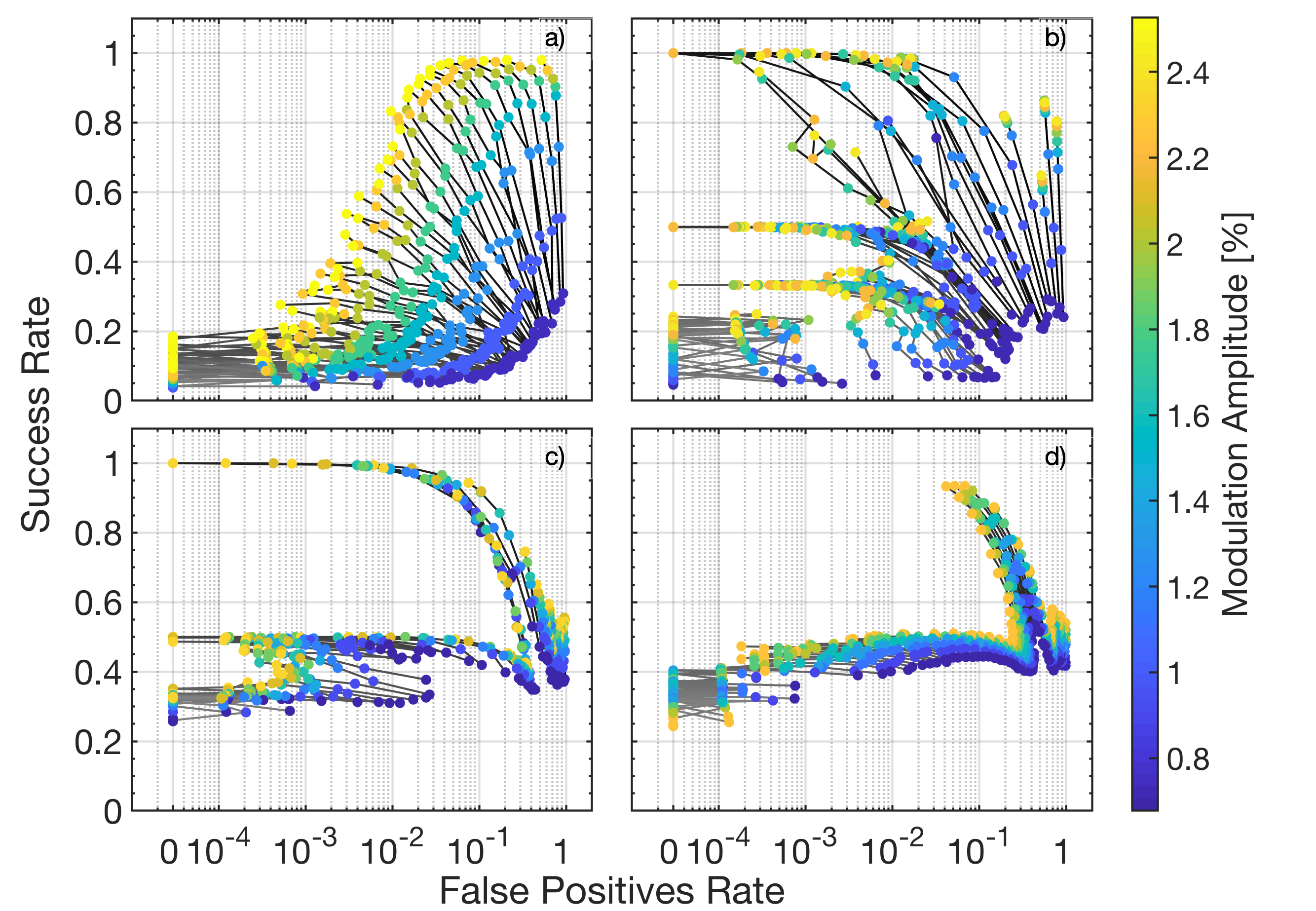}
\end{center}
\caption{ROC curves for (a) $I_{dc}=$ 25~mA, (b) 26~mA, (c) 27~mA, and (d) 28~mA. The lines join points with the same perturbation frequency, while the color scale represents the amplitude in \% of $I_{dc}$. To represent in the logarithmic scale the value FPR$=0$, we have set it to $3\times 10^{-5}$ (labeled as 0 in the x-axis).}
\label{fig:ROC_4panels}
\end{figure}

Finally, in Fig.~\ref{fig:ROC_4panels}, we show  the ROC curves for the four pump currents analyzed before. In each panel we plot the SR and FPR values obtained for all the amplitudes and frequencies studied ($A_{mod}\sim 0.8-2.4$ \% of $I_{dc}$, $f_{mod}=1-80$~MHz). For easy visualization we join the points with iso-frequency lines [as in Fig.~\ref{fig:ROC}(a)] while the color of the points indicate the amplitude of the perturbation. In this way we obtain new insight into the locking transitions.

Figure~\ref{fig:ROC_4panels}(a) clearly shows that, for low pump currents, it is not possible to perfectly entrain the spikes: while the success rate can approach to $1$ (at low modulation frequencies), the number of false positives is always large (revealing many ``natural'', uncontrolled spikes).  At intermediate pump currents [Figs.~\ref{fig:ROC_4panels}(b) and (c)] there is perfect entrainment: the plateaus that were observed in the ISI distributions, Figs.~\ref{fig:ISI_hist_4panels}(b) and (c), are now points lying at SR$=1,{1}/{2}$ and ${1}/{3}$ and FPR$=0$. At the higher laser current perfect entrainment is not seen since the SR approaches, but does not reach, the values $1,{1}/{2}$ or ${1}/{3}$. In addition, the FPR is always large. 

\section{Conclusions}

To summarize, we have studied experimentally the entrainment of a noisy oscillator to a weak external signal and proposed a novel technique, based on the ROC curve, to quantify the degree of entrainment. We have shown that for appropriated parameters it is possible to have full control of the laser output, entraining the optical spikes to the electric perturbations such that the success rate (the number of spikes per perturbation cycle), is equal to $1$, ${1}/{2}$ or ${1}/{3}$, while the false positive rate (the number of spikes which are not triggered by a perturbation), is equal to zero. 
We have also characterized the locking-unlocking transitions, when the perturbation amplitude or frequency vary. The SR and FPR measures unveil different types of transitions: as the amplitude grows, locked behavior is gradually achieved; when the frequency increases, there are rather abrupt locked--unlocked transitions.
Our results demonstrate that the ROC curve is a powerful tool to quantify entrainment, which can be applied to any forced system exhibiting noisy oscillatory behavior.


\section*{Acknowledgments}
This work was supported in part by Spanish MINECO/FEDER (FIS2015-66503-C3-2-P). C. M. also acknowledges partial support from ICREA ACADEMIA, Generalitat de Catalunya.


\end{document}